# Retrospective Evaluation of an Always-on Cherenkov Imaging System for Radiotherapy Quality Improvement

**Authors:**
Daniel A. Alexander MS[1,2‡], Michael Jermyn PhD[1,2], Petr Bruza PhD[1,2], Rongxiao Zhang PhD[1,3,4], Erli Chen MS[5], Savannah M. Decker BS[1], Tatum L. McGlynn BS RTT[4], Rory A. Rosselot BS RTT[4], Jae Lee BA RTT[4], Melanie L. Rose MD[3,4], Benjamin B. Williams PhD[1,3,4], Brian W. Pogue PhD[1,2,3,4], David J. Gladstone ScD[1,3,4], Lesley A. Jarvis MD PhD[1,3,4‡]

[1]Thayer School of Engineering, Dartmouth College, Hanover NH 03755
[2]Dose Optics LLC, Lebanon NH 03766
[3]Geisel School of Medicine, Dartmouth College, Hanover NH 03755
[4]Dartmouth-Hitchcock Medical Center, Lebanon NH 03756
[5]Cheshire Medical Center, Keene NH 03431

[‡]**Corresponding Authors:**
- Lesley A. Jarvis MD, PhD, Section of Radiation Oncology, Department of Medicine, Dartmouth-Hitchcock Medical Center, 1 Medical Center Drive, Lebanon NH 03756. Email: Lesley.A.Jarvis@hitchcock.org.
- Daniel A. Alexander, MS, Thayer School of Engineering, Dartmouth College, 1 Medical Center Drive, Williamson 7, Lebanon NH 03756. Email: Daniel.A.Alexander.TH@dartmouth.edu.

**Conflict of Interest Statement for All Authors:**
DAA, PB, MJ, BWP, and LAJ have competing interests in DoseOptics, LLC.

**Funding Statement:** This work has been funded in part by NIH grant R01 EB023909, with hardware support through R44 CA232879 and the Norris Cotton Cancer Center shared irradiation resources through P30 CA023108.

**Data Availability Statement for this Work:** Research data are stored in an institutional repository and may be shared upon request to the corresponding author.

**Acknowledgements:** The authors would like to thank the entire radiation therapy team at DHMC, specifically Jesse Mayer, Tracy Martin, and Jill Flowers for their assistance with this study.

# ABSTRACT

**Purpose:** Cherenkov imaging is now clinically available to track the course of radiation therapy as a treatment verification tool. The aim of this work was to discover the benefits of always-on Cherenkov images as a novel incident detection and quality improvement system through retrospective review of imaging in our center.

**Methods:** Continuous imaging of all patients was attempted during a 12-month period by automating the acquisition of Cherenkov imaging using an always-on commercial system. Multi-camera systems were installed in two treatment bunkers in the radiation oncology clinic at our center and one bunker in an affiliated satellite clinic. Images were acquired as part of normal treatment procedure and reviewed retrospectively with potential incidents flagged for evaluation by the physician and medical physics teams.

**Results:** In total, 622 patients were imaged as part of this study. In this summary, 9 patients were identified with incidents occurring during their course of treatment that were detected only with Cherenkov imaging. Incidents were found relating to issues during simulation, planning, pre-treatment review, and treatment delivery, however none of the incidents were detected prior to treatment delivery. Primary areas of improvement identified in this study are dose to unintended areas in planning, dose to unintended areas due to positioning, and non-ideal bolus placement during setup. Case studies are presented highlighting the detection of these issues using Cherenkov imaging.

**Conclusions:** All detected events were deemed below the threshold for reporting, but their observation could lead to quality improvement in practice. Perhaps most importantly, the imaging was seamless with no effort required by the radiotherapy team and provided both real-time and permanent records of what was delivered in each fraction.

# 1. INTRODUCTION

Technology advances in radiation therapy (RT) have allowed for higher conformality in dose distributions, allowing for target tumor control with increased sparing of healthy tissues[1]. Necessarily accompanying these advances is a trend toward increased treatment complexity[2]. This has led to the need for robust error mitigation strategies and event reporting in incident learning systems (ILS) for prevention and tracking of incidents in radiation therapy. Adoption of risk mitigation techniques[3] and multilayered detection approaches to incident prevention[4] have informed clinical procedures, such as plan and chart review[5]. Despite these efforts and the safety of radiation therapy as a whole, incidents do occur[6], and human behavior remains one of the leading causes of errors in RT[7]. This study examines the new technology of Cherenkov imaging as a tool for identification and mitigation of some of the delivery errors.

Tighter tolerances on dose delivery have led to the increased adoption of image guided radiation therapy (IGRT), which involves the use of various imaging modalities including MV portal imaging, kV X-ray imaging, cone beam computed tomography (CBCT), magnetic resonance imaging (MRI), and optical surface imaging or surface guided radiation therapy (SGRT)[8–12]. IGRT allows for increased confidence in treatment efficacy, at the expense of increased treatment time and clinic resources. These techniques contribute to reducing incidents during treatment[13], and are used in a majority of clinics[14].

In recent years Cherenkov imaging has emerged as a novel imaging modality in radiation therapy, which is able to visualize the extent of the treatment field and radiation dose on the patient surface[15] due to the nature of the relationship between energy deposition from ionizing radiation and the production of Cherenkov radiation[16]. The utility of this imaging has been demonstrated previously[17–22], and this is bolstered by the lack of additional ionizing radiation or invasive equipment needed to acquire Cherenkov images. It has been estimated that over 40% of reported RT incidents are related to patient setup and positioning[7], and therefore new systems which can serve as a second check of the setup in addition to existing techniques are desirable. Cherenkov imaging has been shown to be sensitive to variations in patient setup in relation to the position of the radiation beam, and therefore is well-positioned to fill this role.

The main purpose of this work was to assess the potential benefits of continuous, always-on Cherenkov imaging for treatment monitoring of a representative patient cohort during one year of clinical operation, to supplement the existing quality assurance program which utilizes a hospital-wide incident learning reporting system that is anonymous, voluntary, and non-punitive. As part of this program, events are reviewed monthly by a multidisciplinary group including radiation therapists, dosimetrists, nurses, physicists, and physicians. The goal of clinic-wide Cherenkov imaging was achieved with the installation of the BeamSite, a newly FDA-cleared medical Cherenkov imaging device for use in radiation therapy treatment monitoring.

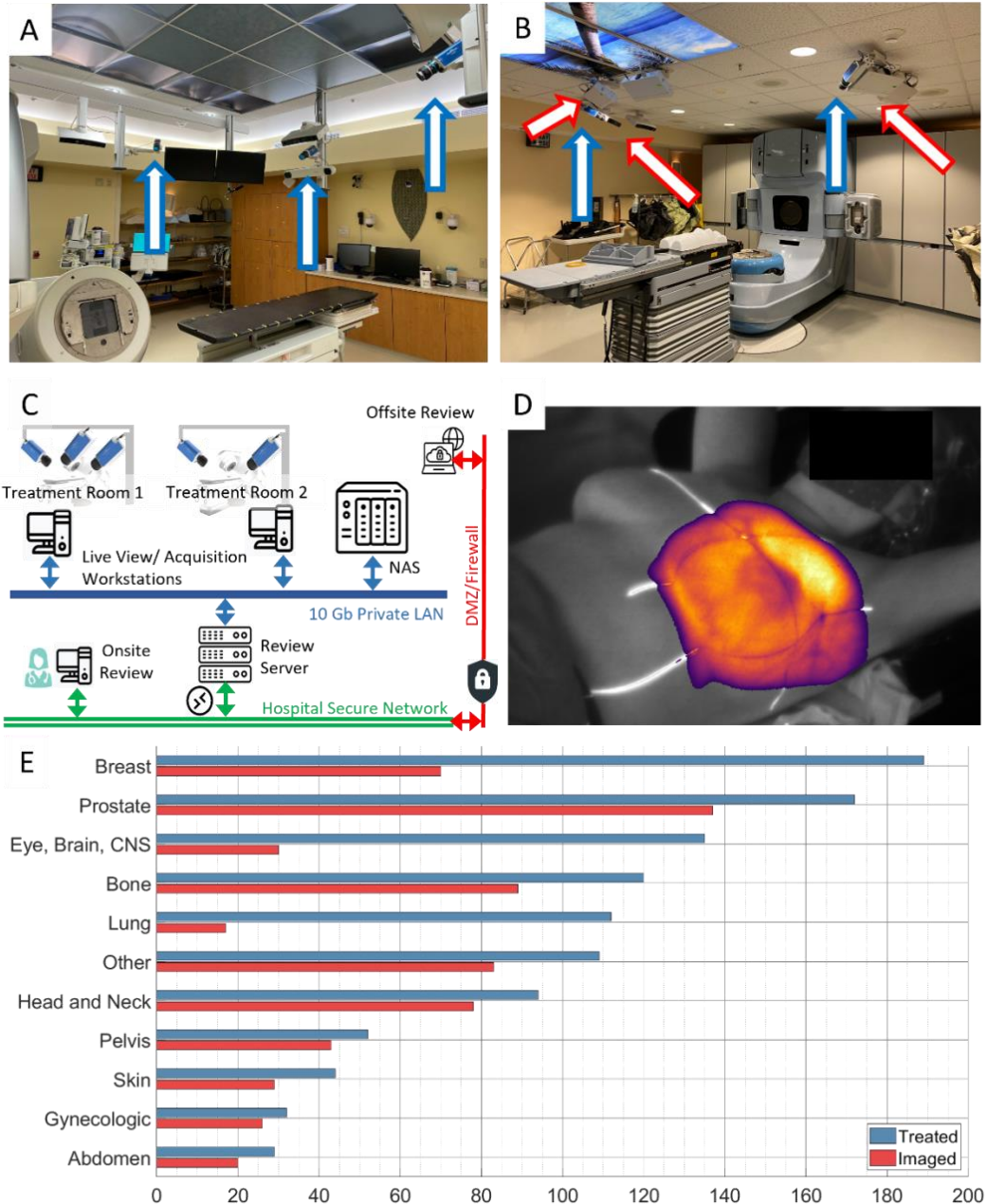

Figure 1. (A) Position of the three permanently mounted BeamSite cameras installed in Bunker A denoted by blue arrows. (B) Position of the two permanently mounted BeamSite cameras installed in Bunker B (blue arrows) and the three custom gated AlignRT pods (red arrows) installed alongside the clinical pods. (C) Schematic of the 10 Gbps dedicated local area network used to transfer high volumes of Cherenkov image data, and the interface with the hospital network. (D) Example Cherenkov image of a left breast treatment generated by BeamSite. (E) Bar graph highlighting the number of patients treated per disease site along with the corresponding number of patients imaged with BeamSite, for the date range defined by the study (October 1, 2020 – October 1, 2021). In total, 622 patients were imaged out of 1088 treated.

## 2. MATERIALS AND METHODS

### 2.1. Imaging Setup

In order to image as many patient treatments as possible, BeamSite cameras (DoseOptics LLC, Lebanon NH) were installed using ceiling mounts in two treatment bunkers at our main campus at the Dartmouth-Hitchcock Medical Center and one treatment bunker at an affiliated satellite clinic at the Cheshire Medical Center. At the main site, three cameras were installed in one bunker containing a Varian TrueBeam as shown in Figure 1A (Varian Medical Systems, Palo Alto, CA), with two cameras positioned laterally and one positioned axially, while two lateral cameras were installed in a second bunker as shown in Figure 1B, containing a Varian Trilogy. At the satellite clinic, two lateral cameras were installed in a bunker containing a Varian TrueBeam. At the main site, both treatment rooms featured surface guidance (AlignRT, Vision RT, London UK), while the satellite clinic did not contain an SGRT system.

The cameras were outfitted with a 50 mm f/1.2 lens with a tilted focal plane, which was aligned to match a typical patient surface more closely. Additionally, optical filters were mounted on each lens to reject specific light sources in the room, including green alignment lasers and infrared transceivers on the gantry. Broadband light sources such as the field light projector and the optical distance indicator (ODI) were not filtered, and therefore therapists were asked to mute them, when possible, in order to preserve the fidelity of the images. Each camera was connected to a workstation running the BeamSite software in the corresponding console area using a hybrid copper-fiber USB cable, allowing for both power transmission and high data throughput. Images were acquired during treatment at 20 frames per second, and data was saved in 16-bit RAW format in a rolling storage buffer on each local workstation, with a capacity of 3-5 days' worth of imaging data depending on the camera configuration.

The camera image intensifiers were gated to the linac pluses using an internal radiation trigger unit (RTU) in order to capture the emitted Cherenkov light from irradiated patient tissue with a reasonable signal-to-noise ratio (SNR) with the rooms lights on[23]. Additionally, this signal informed the software of the beam-on status, which allowed for data to only be saved when treatment was occurring, while live images were displayed on the monitor to the therapy team continuously. No patient-specific information was entered into the software for each treatment, and instead image stacks were saved in chunks bounded by beam-off times above a specific duration threshold, fixed at two minutes. This threshold was purposefully set at a conservatively low value in order to avoid multiple patients' treatments being captured in one acquisition, at the expense of the possibility of one treatment being split into two or more acquisitions if enough time passed between beams (Supplementary Figure 1).

### 2.2. Data Transfer and Storage

To facilitate efficient data transfer between each console workstation and central storage, a dedicated 10 Gbps local area network was established at both the primary and satellite clinics to avoid the transfer of large amounts of data over the hospital network, as depicted in Figure 1C. Data was copied nightly using an automated script from each console workstation to the central network attached storage (NAS) server at each site with an available capacity of over 200 TB per linac. A custom query to the Aria database also ran daily to produce a file containing details for each treatment, including machine ID, patient medical record number (MRN), treatment description, and treatment start and end timestamps. These files were parsed automatically, and patient metadata was assigned to the appropriate Cherenkov image acquisition by cross-referencing the timestamps from Aria and BeamSite. This process allowed for data to be accessed and reviewed by clinicians in a patient-centric workflow without any manual input from the treatment team. All metadata was written to a database file created by a custom version of C-Dose Research (DoseOptics LLC, Lebanon, NH), which enabled advanced search through all acquired patient images.

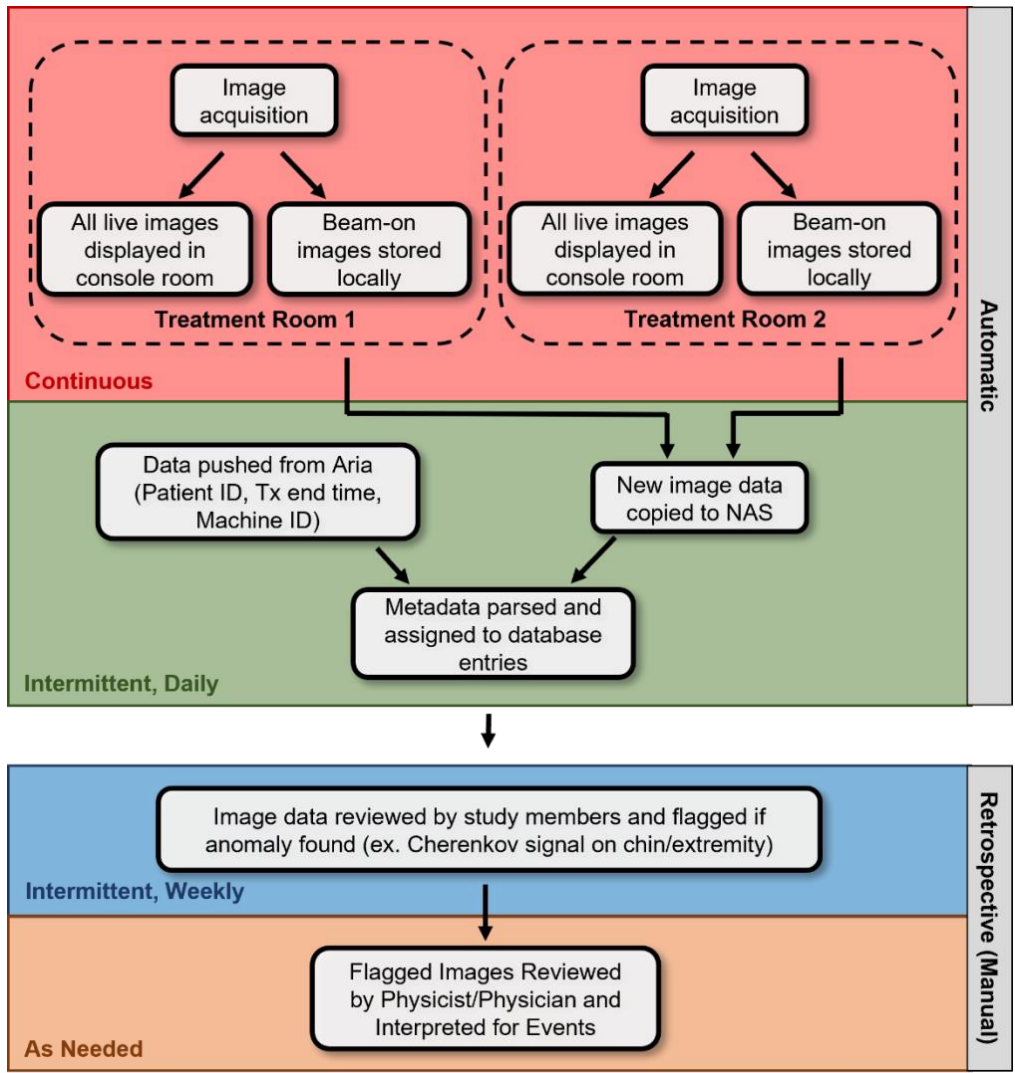

Figure 2. Flow chart highlighting the Cherenkov image data pipeline from acquisition to review. Images are acquired, transferred, and processed automatically as shown in the red and green boxes. Subsequently, image review, flagging, and interpretation for events occur manually on a retrospective basis.

### 2.3. Data Review

Cherenkov image data was reviewed as part of an IRB-approved clinical study, with a date range of October 1, 2020 to October 1, 2021. Cherenkov imaging began at the main site on November 11, 2020, and at the satellite location on March 4, 2021. During this time period, cameras experienced some downtime for upgrades, repair, or software debugging. Live data was displayed in real time on a monitor in each treatment console, allowing for therapists to monitor the beam position on the patient, although training on interpretation of the images was minimal during this initial phase of clinical imaging. All image data was systematically reviewed retrospectively by study team members for detectable anomalies. When found, anomalous datasets were flagged for further review by a medical physicist or radiation oncologist to interpret the impact of the finding. This process is outlined in Figure 2.

### 2.4. Compatibility with Surface Guidance Devices

The Cherenkov imaging system was not natively compatible with AlignRT surface monitoring, so treatments that employed the use of surface guidance in the two treatment bunkers at the main campus were unable to be imaged. These treatments included frameless SRS and SBRT (primarily in the TrueBeam bunker) and breast (primarily in the Trilogy bunker). In order to facilitate breast imaging, a custom set of three gated AlignRT pods were installed in the treatment bunker containing the Trilogy alongside the standard three pods as shown with the red arrows in Figure 1B. These pods were installed partway through the study, in May of 2021. The custom pods were time-gated to the RTU signal produced by the cameras, which temporarily disabled the projectors during linac pulses. This customization did not alter the standard pods, which were still used for all surface monitoring and capture and allowed the standard AlignRT software to be used. A characterization dataset was acquired for the customized system, and due to the low duty cycle of the linac pulses, this customization introduced no alteration in performance of the AlignRT system. This solution was not implemented in the second bunker due to the limited availability of custom AlignRT units.

## 3. RESULTS

### 3.1. Cohort Results

In total, 622 patients were imaged as part of this study. A breakdown of patients imaged and treated during the study date range per disease site is shown in Figure 1E. Patients whose images were acquired but were uninterpretable due to the use of surface monitoring, primarily including patients treated with frameless SRS, lung patients treated with SBRT, and some whole breast irradiation patients, were excluded from this total. This is highlighted in the discrepancy in Figure 1E between imaged and treated patients, specifically for disease sites like breast, lung, and brain. To assess the results of the retrospective review, the following definition of 'incident' from Ford et al[6] was used: "An unwanted or unexpected change from normal system behavior which has the potential to cause an adverse effect to persons or equipment." Out of this cohort, 9 (1.4%) patients were found to have an incident occur during their treatment that was caught only with Cherenkov imaging. Using non-mutually exclusive categorizations adopted from Hallaq et al[24], of these 9 events, 1 corresponded to an issue during simulation, 2 corresponded to issues during planning, 3 corresponded to issues during pre-treatment review, and 6 corresponded to issues during treatment.

### 3.2. Case Studies

#### 1.1.1. Detection of Sub-Optimal Planning

Two cases of sub-optimal planning were revealed by review of Cherenkov images. Case 1 involved a 4-fraction boost treatment to the left breast. It was noticed by the physicist that there was Cherenkov signal emitted from the contralateral breast, and visualization of the surface dose rendered on the patient surface revealed that dose was deposited to that area as planned, as displayed in Supplementary Figure 2. Review of the treatment plan highlighted dose to the contralateral breast at the 10% isodose line, which was not appreciated during plan approval.

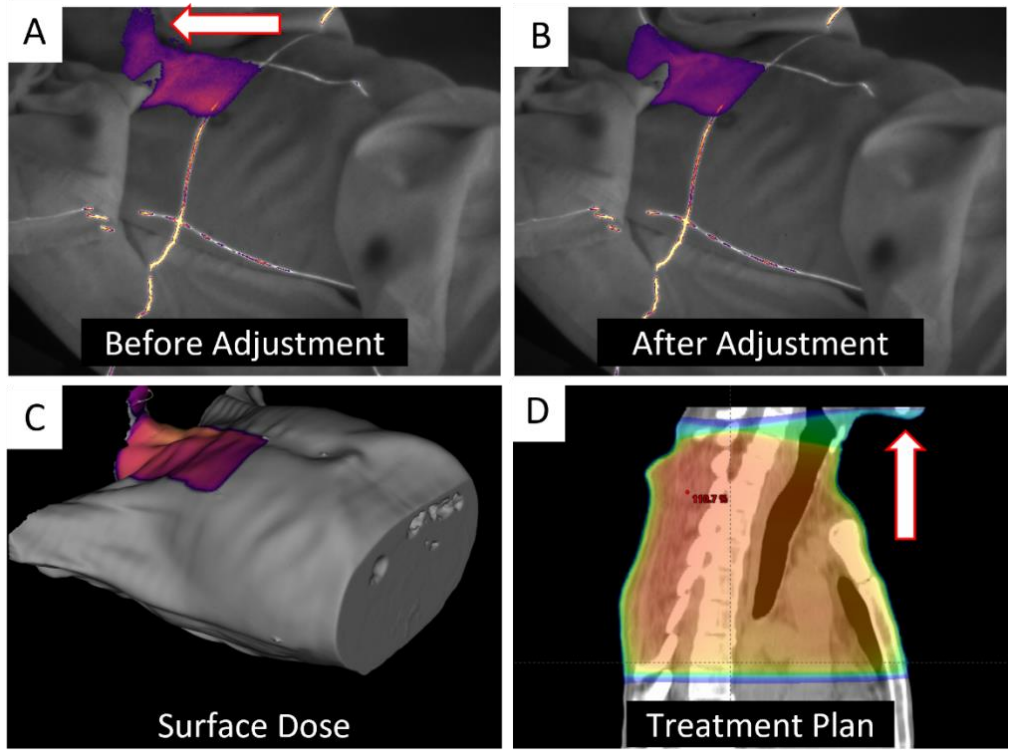

Figure 3. Case 2 showing dose to the patient's chin during a palliative thoracic spine treatment, which was noticed live by a radiation therapist, prompting intervention to position the chin slightly further back. Cherenkov image frames accumulated from the beginning of treatment to beam interruption are shown in (A), while Cherenkov image frames accumulated from beam continuation to the end of the relevant field are shown in (B). The surface dose from the relevant field as extracted from the TPS projected on the patient surface extracted from the simulation CT scan is shown in (C), visualized from the camera perspective. The dose to the chin as shown in (C) was verified by checking the treatment plan (D).

Case 2 involved a 3-field palliative thoracic spine treatment, where after setup and delivery of the first PA field Cherenkov signal was observed from the patient's chin by the radiation therapist during treatment as shown in Figure 3A. This led to the therapy team stopping the treatment, entering the bunker, and readjusting the patient's chin past the superior edge of the exit field. Upon resuming the treatment, the therapists noted that the beam was no longer intersecting the chin (Figure 3B). Upon retrospective review, both the surface dose render and a sagittal view of the treatment plan showed that due to the patient position during simulation, exit dose was indeed delivered to the chin as planned (Figure 3C,D). Additionally, this highlighted a sub-optimal simulation CT scan, as the chin was just shy of the superior limit of the scan.

*1.1.2. Dose to Unintended Area Due to Imperfect Positioning*

Three cases of dose being delivered to unplanned healthy tissues were found during retrospective review (Cases 3-5). Case 3 involved a 3-field conformal treatment to the sacrum for 10 fractions. Upon review of the images post-treatment, it was found that the patient's left arm was positioned in the exit beam of the RPO field, as shown in Figure 4B, for 7 out of the 10 treatment fractions. This was in contrast to the proper left arm positioning, which is shown in Figure 4A. The physicist estimated the radiation dose delivered to the arm to be 4.5 Gy total from these 7 instances. All imaged treatment fractions for this patient are displayed in Supplementary Figure 3, while a video feed of fraction 6 is shown in Supplementary Video 1 (10.6084/m9.figshare.16811308). Case 4 involved a similar 3-field conformal treatment to the lumbar spine delivered in 5 fractions, for which the patient's left arm was found to have moved into the exit beam during irradiation with the RPO field during the first fraction only (Supplementary Video 2: 10.6084/m9.figshare.16811326).

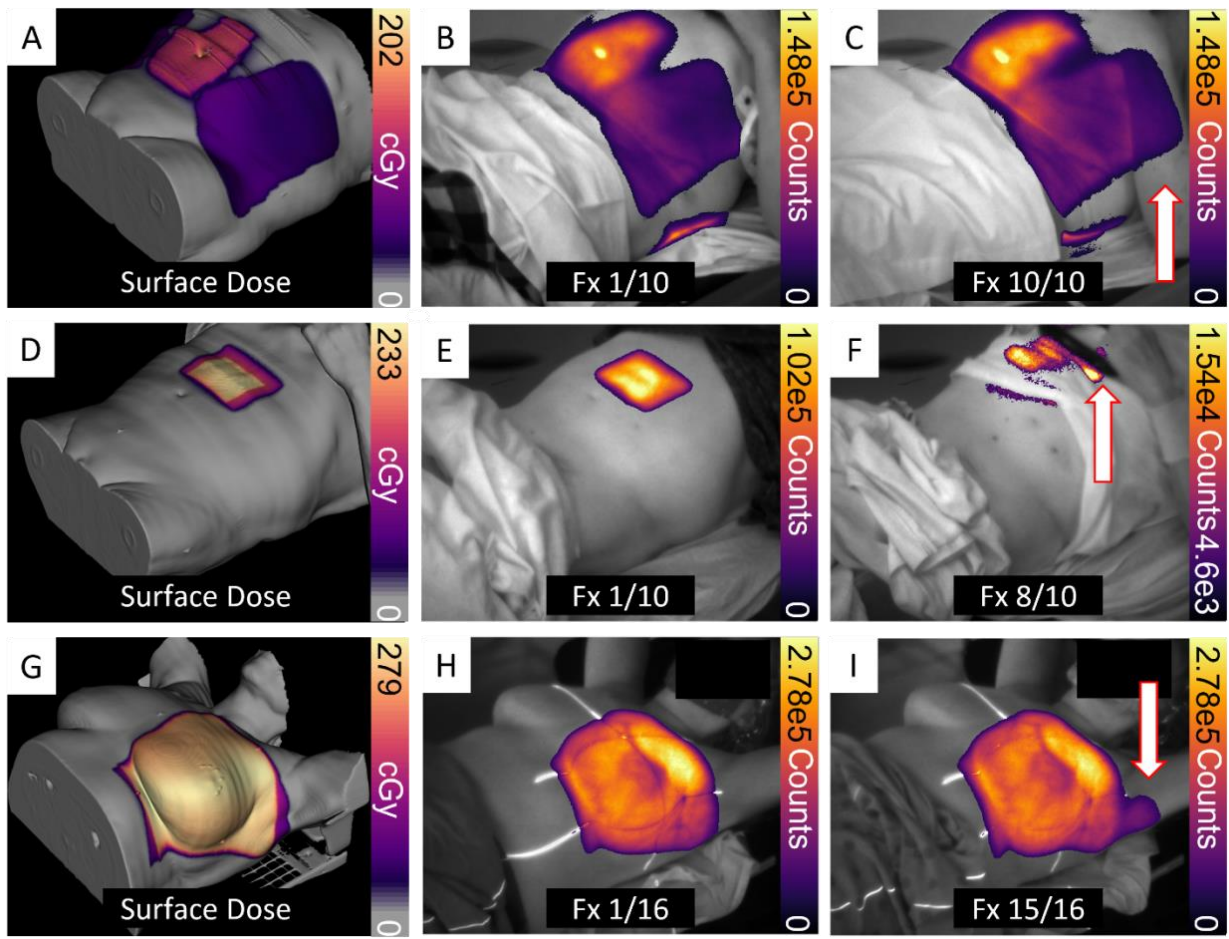

Figure 4. Cases 3, 5, and 6 showing dose to unintended areas. Case 3 (A-C) shows improper arm placement (7/10 fractions) leading to 4.5 Gy exposure from an exit beam. Case 5 (D-F) shows non-ideal hand position leading to minor exposure (1/10 fractions). Case 6 (G-I) highlights extraneous dose to the left armpit from small deviation in left arm position (1/16 fractions). All surface dose renderings were extracted from the treatment plan and overlaid on the patient CT surface, while red arrows highlight event regions.

Case 5 consisted of a palliative spine treatment which intersected the patient's left hand in 1 out of 10 fractions, as shown in the comparison between Figures 4E and 4F. Similar to Case 3, this was a result of imperfect hand positioning during setup. This case also highlights the inconsistency of the patient's clothing position, which impacts the visibility of the Cherenkov emission on the patient surface (Supplementary Figure 4). Case 6 involved a 16-fraction left-sided whole breast radiation therapy treatment, with two sets of tangents using 6MV and 10MV beams, respectively. It was found in retrospective review that during fraction 15, a slight change in the position of the patient's left arm led to extraneous dose to the armpit from the exit beam of the 6MV and 10MV RAO fields (Supplementary Video 3: 10.6084/m9.figshare.16811335) as shown in Figure 4I, in contrast to the proper arm placement shown in Figure 4H. All the imaged fractions for Case 6 are shown in Supplementary Figure 5.

*1.1.3. Improper Bolus Placement*

Improper bolus placement was highlighted in three cases (Cases 7-9) leading to uncovered irradiated surfaces on the patient and potential underdoses. All three of these cases included post-mastectomy irradiation of the chest wall with bolus coverage. Case 7, shown in Figure 5A-C, involved a left chest wall treatment where the entrance and exit dose from the LPO tangent was planned to be covered with bolus. Variation in bolus placement day-to-day led to the medial chest wall being uncovered for 7 imaged fractions. Additionally, it was found that two adjacent pieces of bolus was need for adequate bolus coverage, however the second piece was not used in these 7 fractions (Supplementary Figure 6). Cases 8 and 9 involved right chest wall treatments with bolus coverage as shown in Figure 5D-F (Case 8). The bolus placement left portions of the lateral entrance field uncovered for 3 imaged fractions in Case 8 (Supplementary Figure 7) and 13 imaged fractions in case 9.

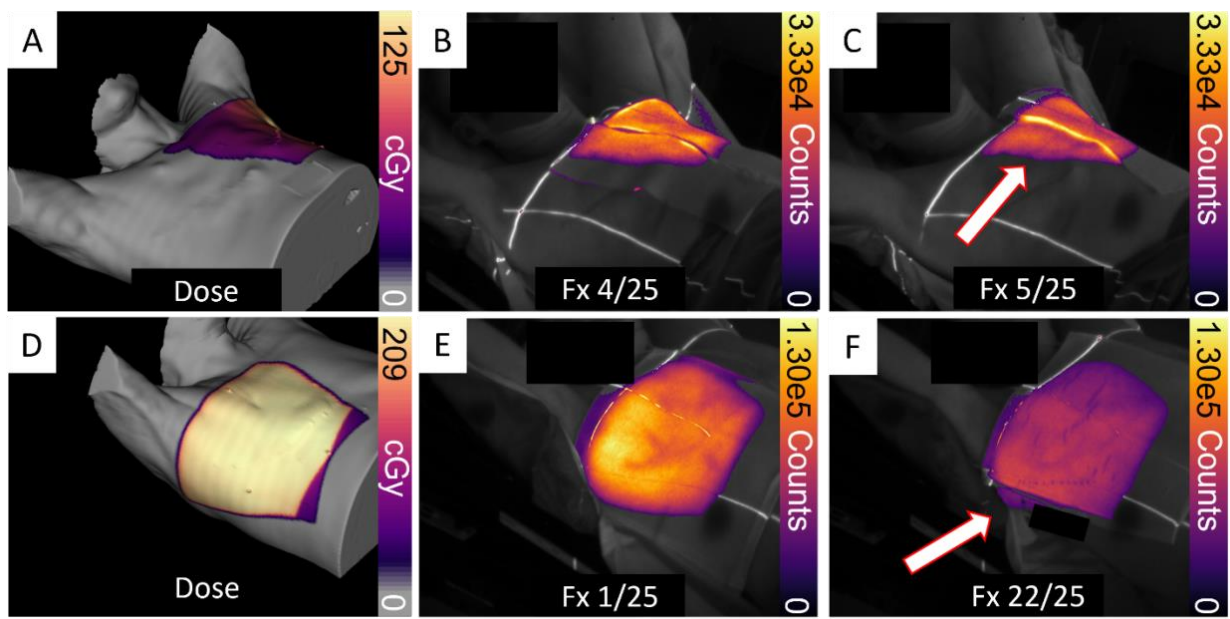

Figure 5. Cases 7 and 8 showing imperfect bolus coverage leading to uncovered irradiated regions of chest wall. Case 7 shows proper coverage using two adjacent pieces of bolus on fraction 4 (B), and insufficient coverage leading to potential underdose from the exit side on fraction 5 (C). Panels A-C show surface dose and Cherenkov images from only the relevant LPO field. Case 8 (D-F) shows a similar but less extreme bolus placement issue on fraction 22 (F) compared with proper coverage shown on fraction 1 (E). Panels D-F

correspond to the entire treatment fraction. All surface dose renderings were extracted from the treatment plan and overlaid on the patient CT surface, while red arrows highlight event regions.

## 4. DISCUSSION

Table 1 displays categorical events that are detectable with Cherenkov imaging, with comparison to existing technologies, and has been informed by the cases resulting from this study. All incidents detected as part of this study represented minor impacts to the treatment, however they all represent opportunities for quality improvement. The Cherenkov imaging system served as a secondary observation tool and was able to catch issues with bolus placement consistency, patient positioning, and undesirable treatment planning choices. Since the Cherenkov video feed was available for the therapists to view during treatment but live viewing was not specifically requested, only one incident was found live and responded to during treatment. It is important to note that most therapists were not trained in use of the system during this early phase of its installation. However, all events were detectable live during treatment, and could have been responded to if the Cherenkov video feed were monitored by the therapists live. This underlines one limitation of this study, being that with the manual review of over 8,000 Cherenkov image datasets on a retrospective basis it is possible that certain events may have been missed by the study team, leading to a potential underreporting of the number of incidents found using this system. In an ideal implementation of this system, live viewing of the Cherenkov images as the treatment is occurring would serve as the most robust method for catching potential delivery issues. Additionally, future work will involve the development and validation of automated image analysis methods, which may serve as the second step to catch delivery incidents that are not found by the therapy team live.

In addition to the points made in the Table 1, one notable benefit of this system compared with both comparable technologies and previous clinical implementations of Cherenkov imaging is that the only interaction with the system is passive observation, with no need for manual operation of any kind. Patient information is matched to images automatically, and images are always being acquired and saved when the beam is on. However, the robustness of this automated process requires additional validation to characterize potential failure modes.

As a camera system capable of imaging the background treatment scene, Cherenkov imaging provides many of the benefits of CCTV camera systems which are installed in all treatment bunkers, however feedback from therapists during this study indicated that the field of view and camera perspective offered by the Cherenkov cameras provide superior patient visualization than standard CCTV cameras. Additionally, SGRT and portal dosimetry are perhaps the most comparable existing systems to Cherenkov imaging, and both have advantages and disadvantages to this technology. In reality Cherenkov imaging presents as a novel imaging modality that is best used in a supplementary manner to existing systems, and this study highlights adverse treatment scenarios that were not detected using the multilayered "swiss cheese model" approach to incident prevention. No other IGRT modality is able to visualize a wide field of view showing the patient surface along with the position of the radiation beam, and this combination provides a powerful way to increase confidence in treatment setup.

Table 1. Event categories detectable by Cherenkov imaging compared with existing systems.

| Event Category | Event Subcategory | Existing Systems | Cherenkov Imaging |
|---|---|---|---|
| Positioning/Setup | Accessory Placement | Setup Photos<br>External Markings | • Can visualize bolus position relative to entrance and exit beam<br>• Can only see beam position once treatment starts |
| | Positioning Inside Treatment Area | Portal X-ray<br>Orthogonal X-ray<br>CBCT<br>MRI<br>RF Tracking<br>Setup Photos<br>SGRT<br>External Markings<br>Field Light | • Verifies position visually, from intuitive perspective<br>• Not sensitive to internal anatomy<br>• No ionizing radiation<br>• Impeded by opaque objects<br>• Sensitive to anatomical features (surface vessels, scars, etc.)<br>• Can only see beam position once treatment starts<br>• Sensitive to exit beam position |
| | Positioning Outside Treatment Area | Setup Photos<br>SGRT<br>External Markings | • ROI is only limited by field of view<br>• Final setup is recorded during treatment<br>• Can only see beam position once treatment starts |
| Treatment Monitoring | Patient Position Compliance | CCTV Cameras<br>MRI<br>SGRT | • Close-up FOV of patient<br>• Camera viewpoint of patient surface<br>• ROI is only limited by field of view<br>• Real-time<br>• No quantitative positional information |
| | Patient Motion | CCTV Cameras<br>MRI<br>SGRT | • Visualize respiratory motion (breathing, coughing)<br>• Visualize Limb/head motion<br>• No quantitative positional information |
| | Dose Area | In vivo point dosimeters<br>Portal dosimetry | • Provides 2D intensity map of irradiated region and surrounding area<br>• No quantitative dose measurements<br>• Independent system from linac |
| | Field Shape | Portal dosimetry<br>Field Light | • Independent system from linac<br>• Sensitive to both entrance and exit beam shape |

The value of Cherenkov imaging various with different treatment modalities, which is highlighted by the case studies presented earlier in this paper. All treatments for which incidents were found involved medium to large, mostly static fields, which skews the benefits of Cherenkov imaging towards non-IMRT treatments. While Cherenkov imaging does provide visualization of small beamlets delivered as part of IMRT treatment plans, current imaging technology is limited by the intrinsically low signal-to-noise ratio of these beamlets, geometrical occlusion at certain gantry angles, and the challenging interpretability of cumulative surface dose maps for highly dynamic treatments. Additionally, the technical incompatibilities of Cherenkov imaging with some SGRT monitoring systems presented a barrier for imaging many SRS, SBRT, and breast treatments, although the custom solution implemented in the second half of the study reduced the impact this had on the number of breast patients imaged. Future SGRT systems could afford compatibility with Cherenkov imaging by implementing a similar solution. We also found that clothing placement, especially for prostate patients which comprised over 20% of the imaged patient cohort, presented a significant challenge in the interpretation of the images, as it blocked optical emission from the patient surface from reaching the camera and limits the extent of the observable field. This challenge could be overcome in a clinical deployment of always-on Cherenkov imaging with adequate training of radiation therapist users.

## 5. CONCLUSIONS

Always on Cherenkov imaging as implemented in this study demonstrated the ability to detect issues with treatment delivery that were not prevented using the standard mitigation and QA strategies employed in our clinic and served as a final check at the moment of treatment delivery. This study reported an implementation of an always-on Cherenkov imaging system as a quality improvement tool, wherein 622 patients were imaged from October 2020 to October 2021. In this cohort, 9 (1.4%) patients had an incident identified using Cherenkov images which were not otherwise identified with any other system. Future work will involve the expansion of use of this technology to other hospitals and health systems, in order to better characterize the role of Cherenkov imaging in the context of RT quality control.

**SUPPLEMENTARY INFORMATION**

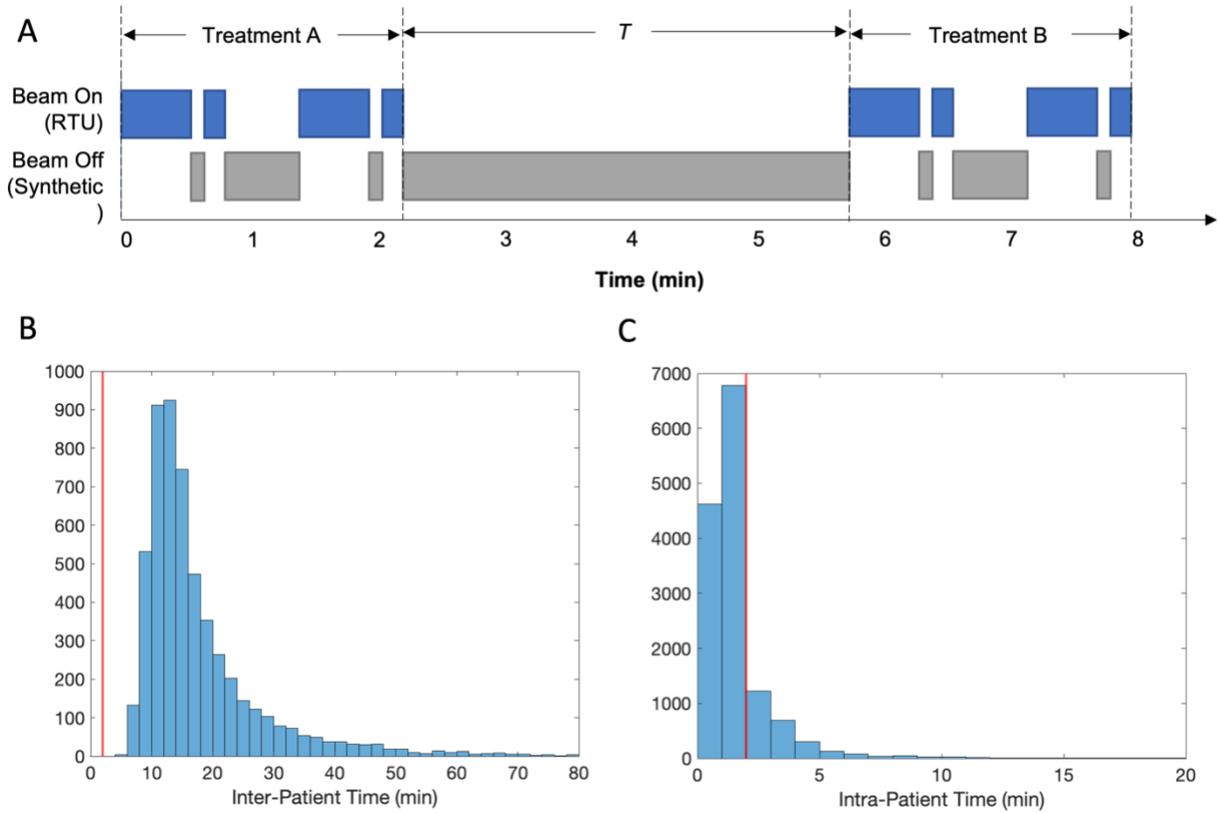

Suppl. Figure 1. (A) Temporal schematic of storage of Cherenkov image chunks. A threshold *T* is chosen such that beam off time in excess of *T* minutes will initiate a new chunk. (B) Histogram of inter-patient beam-off times, with the chosen value of T = 2 min shown as a red line. (C) Histogram of intra-patient beam-off times, with the chosen value of T = 2 min shown as a red line.

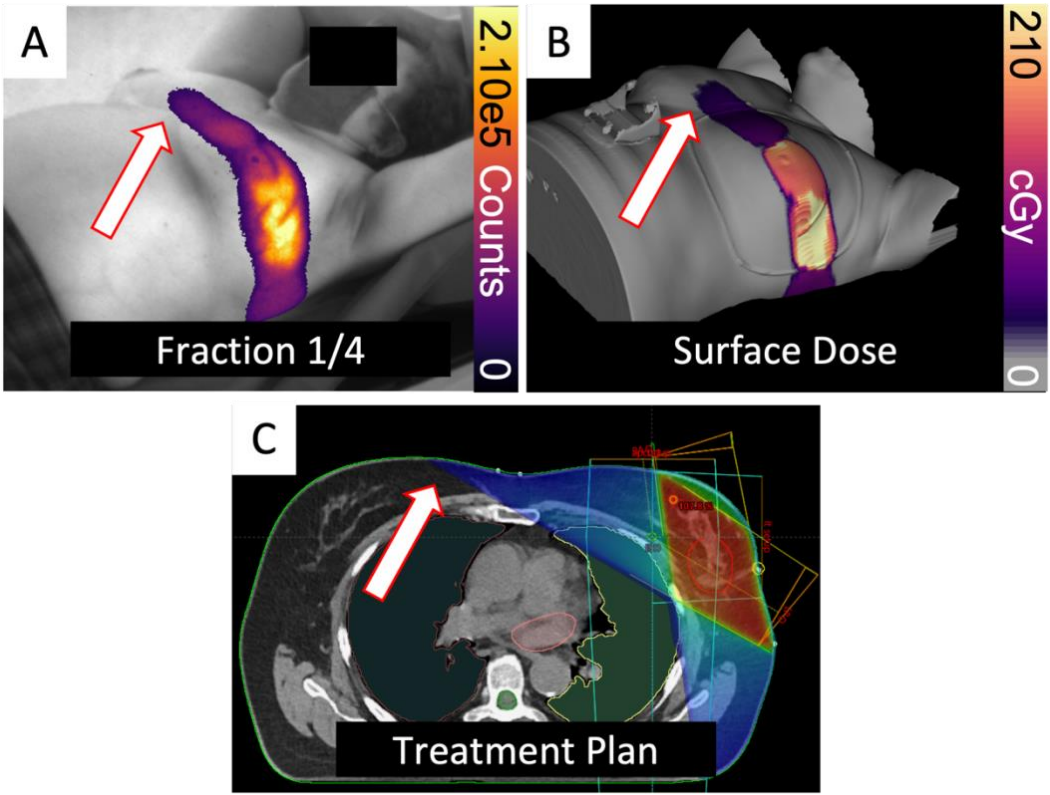

Supplementary Figure 2. Case 1 showing dose to the patient's contralateral breast during a four-fraction boost treatment to the left breast. Signal was observed on the contralateral breast in the Cherenkov video feed (A) indicted with a red arrow, prompting review of the plan by the physicist. Surface dose visualized on the patient CT surface shows dose at the contralateral breast as planned (B), which was confirmed in the TPS by decreasing the visualized dose threshold to 10% of the prescription.

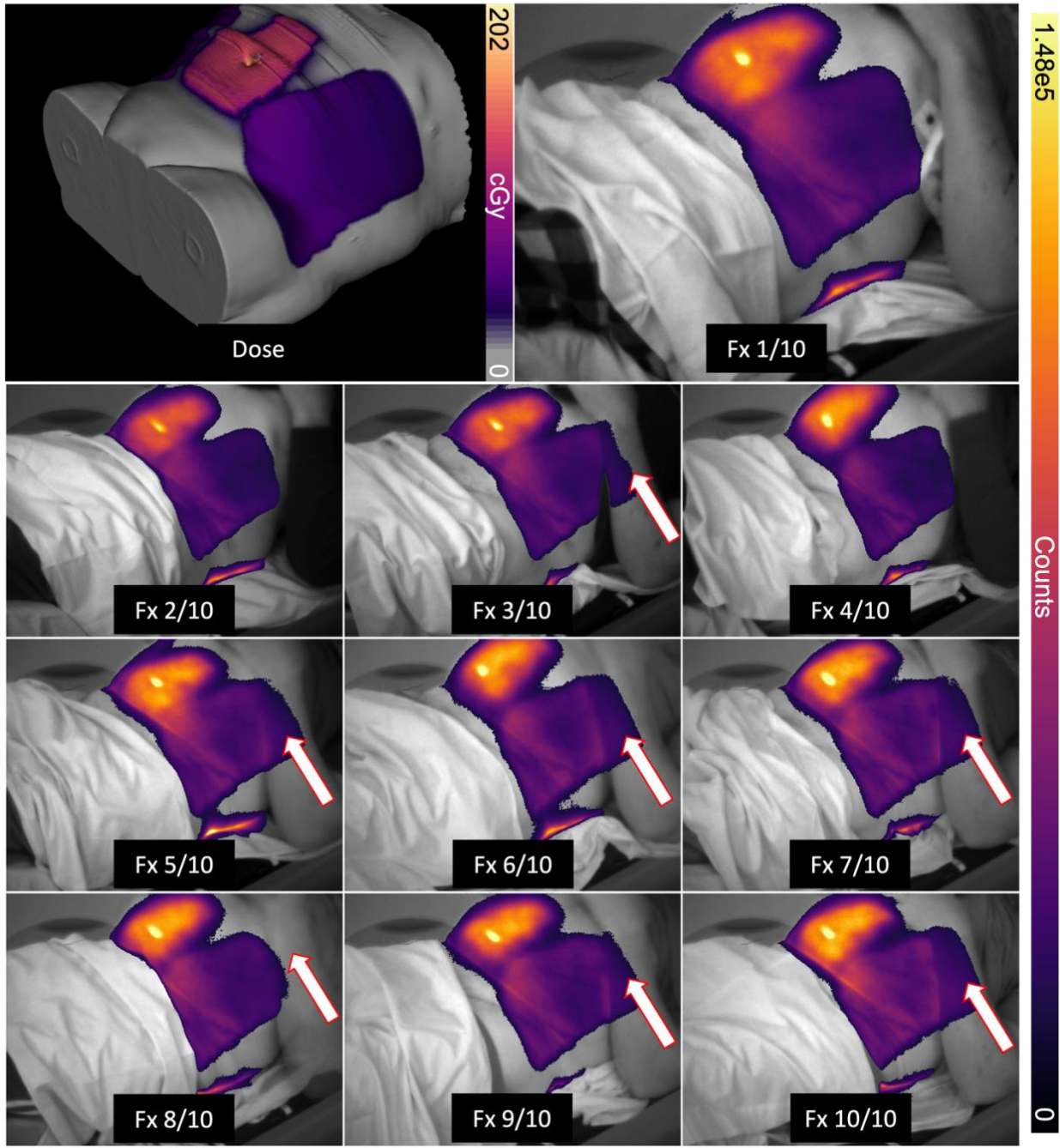

Supplementary Figure 3. Mosaic of all imaged fractions for Case 3, showing exit dose to the patient's left arm in fractions 3 and 5-10. Event regions are indicated with red arrows.

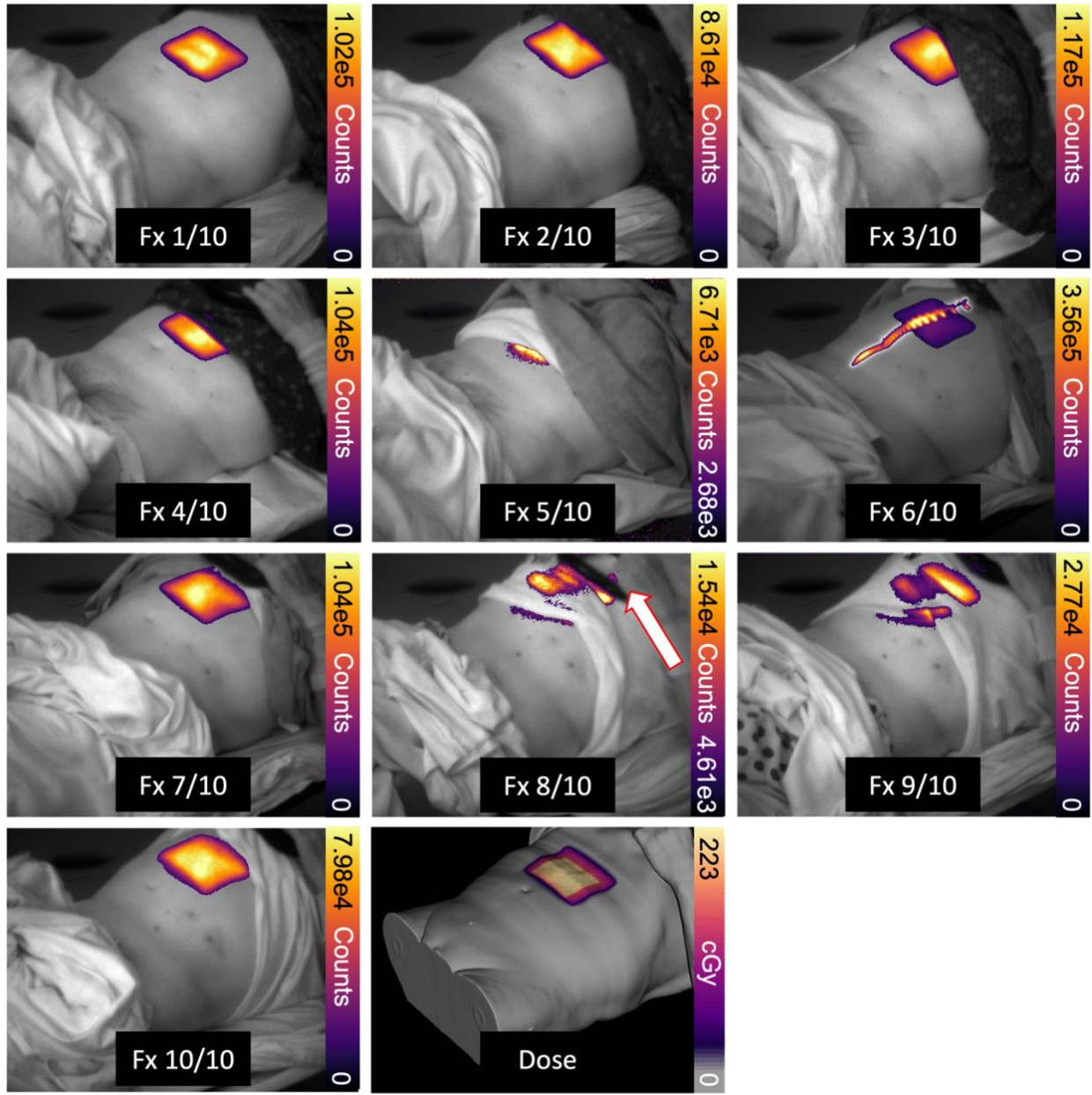

Supplementary Figure 4. Mosaic of all imaged fractions for Case 5, showing exit dose to the patient's left hand in fraction 8, indicated with a red arrow.

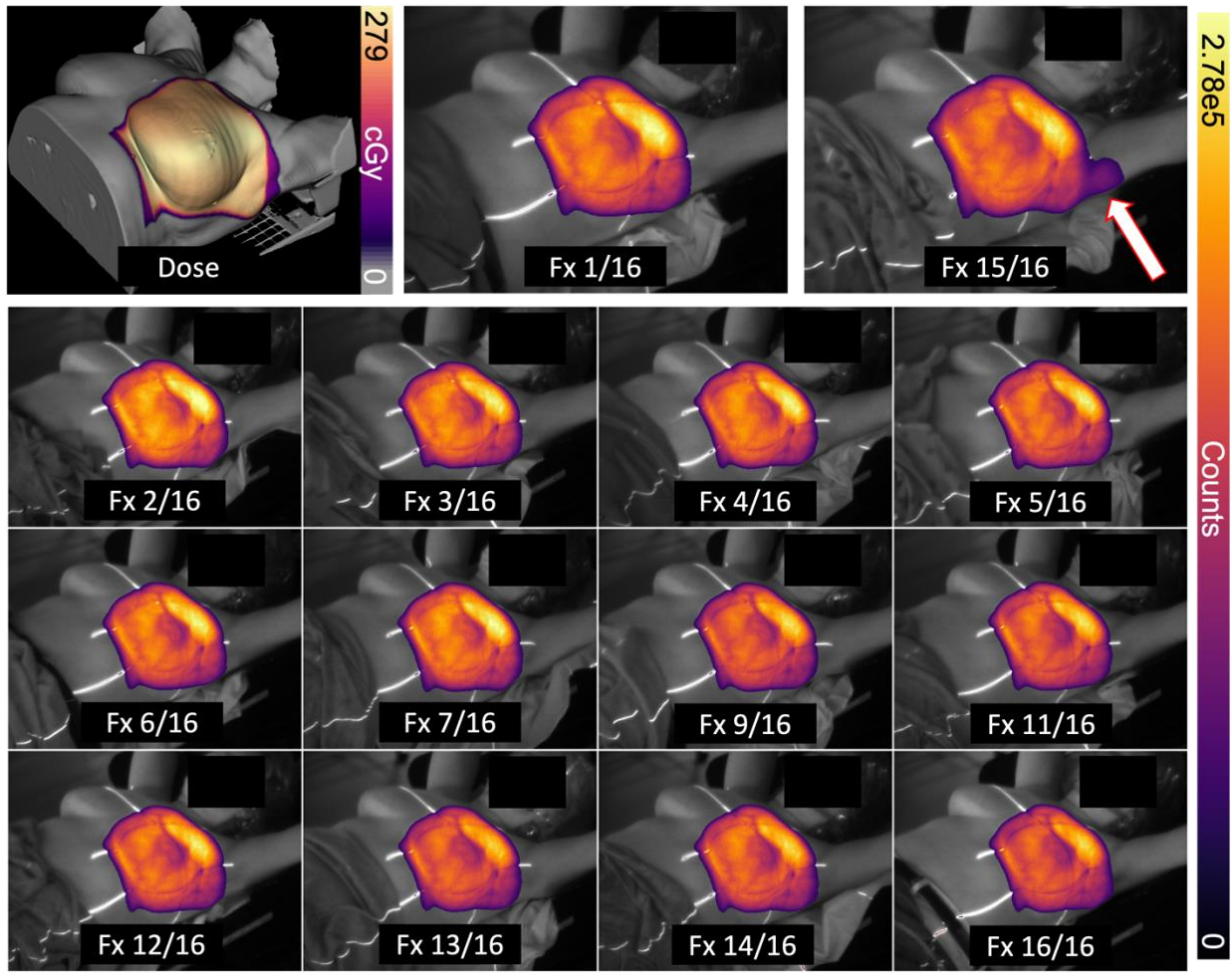

Supplementary Figure 5. Mosaic of all imaged fractions for Case 6, showing extraneous exit dose to the patient's left armpit hand in fraction 15, indicated with a red arrow.

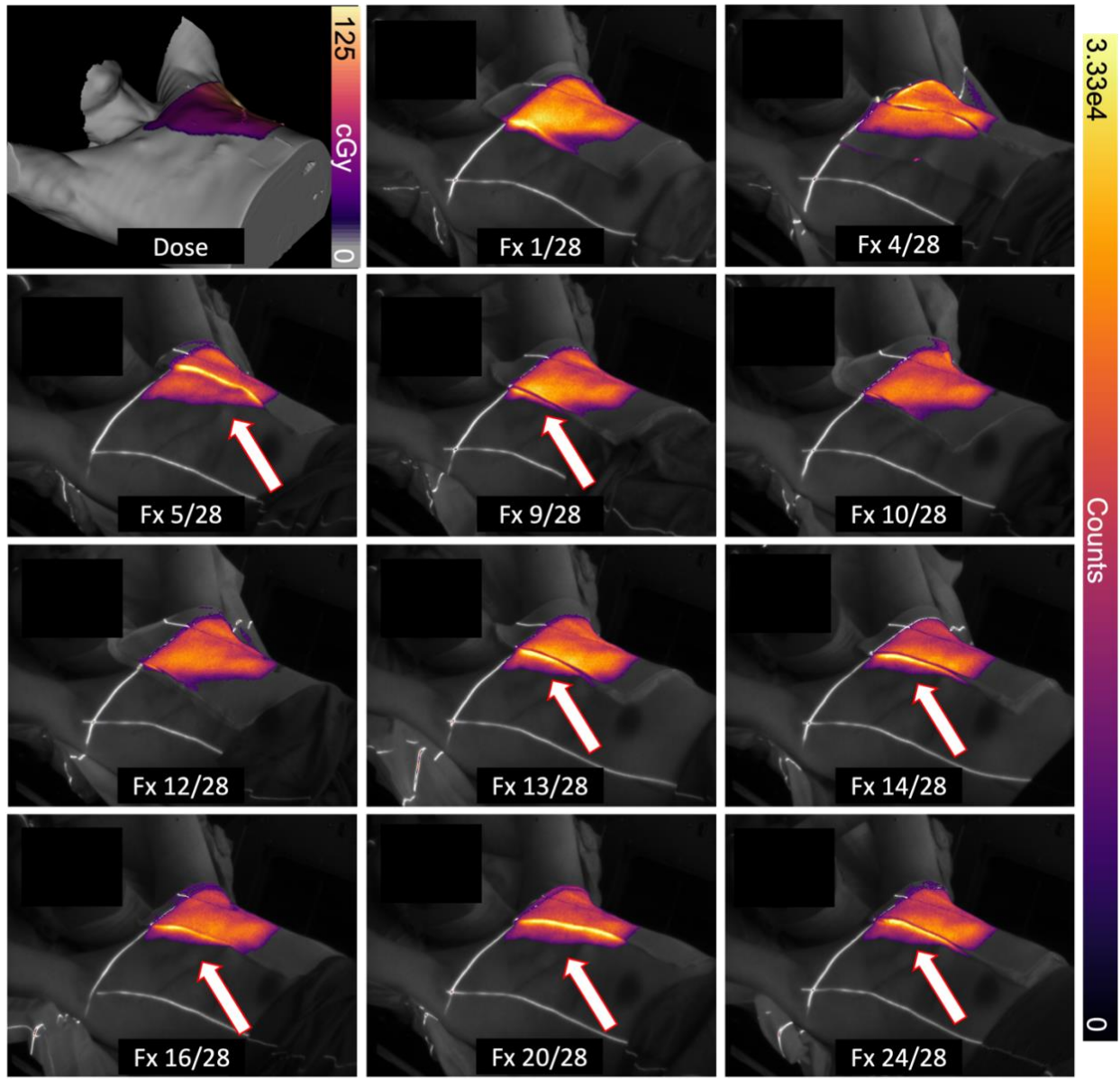

Supplementary Figure 6. Mosaic of all imaged fractions for Case 7, showing exit dose to the patient's medial chest wall not covered by bolus as intended, in fractions 5, 9, 13, 14, 16, 20, and 24. Event regions are indicated with red arrows. Missing fraction images are due to untimely camera downtime.

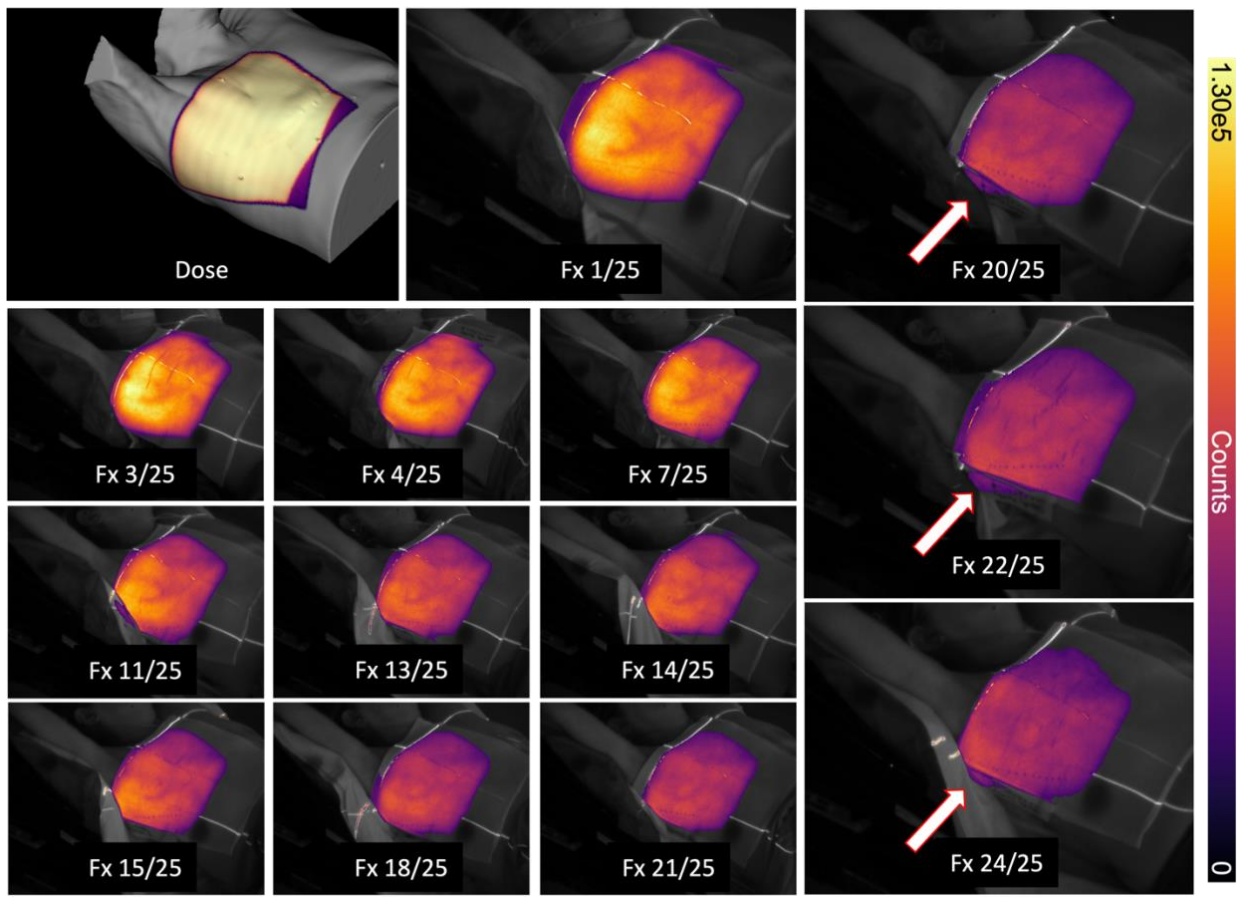

Supplementary Figure 7. Mosaic of all imaged fractions for Case 8, showing entrance dose to the patient's lateral chest wall not covered by bolus as intended, in fractions 20, 22, and 24. Event regions are indicated with red arrows.